\documentclass[prd,aps,showpacs,preprintnumbers,amssymb]{revtex4}
\usepackage{axodraw}
\usepackage{color}
\usepackage{epsf}

\def\e3p{$\eta \rightarrow 3 \pi$}

\begin{document}
\title{%
\hfill{\normalsize\vbox{%
\hbox{}
 }}\\
{ $\Phi^4$ theory is trivial}}

\author{Renata Jora
$^{\it \bf a}$~\footnote[2]{Email:
 rjora@theory.nipne.ro}}

\affiliation{$^{\bf \it a}$ National Institute of Physics and Nuclear Engineering PO Box MG-6, Bucharest-Magurele, Romania}

\date{\today}

\begin{abstract}
We prove that the $\Phi^4$ theory is trivial for any values of the bare coupling constant $\lambda$  thus extending previous results referring to very strong couplings to the full range of values for this parameter.  The method is based on all order properties of the partition and two point functions in the path integral formalism.
\end{abstract}
\maketitle

\section{Introduction}

Perturbative study of many quantum field models has lead to many impressive results: beta functions for QED and for the $\Phi^4$ theory are known  up to the fifth order whereas that of QCD up to the fourth order \cite{Vladimirov}-\cite{Baikov}.  These studies however are based on the expansion in the small coupling constant such that the behavior at large couplings is completely missed.  Although the possibility of a perturbative  expansion in the strong coupling constant has been explored  over the years \cite{Kovesi}-\cite{Bender3} no definite method emerged.

Of significant importance is the $\Phi^4$ theory at strong coupling whose triviality \cite{Wilson}-\cite{Jora} has been the subject of intensive debate. In essence the triviality of the $\Phi^4$ theory means that the renormalized  coupling constant $\lambda_R$ vanishes in the limit of large cut-off and the model behaves  like a non-interacting field theory. Many previous theoretical studies suggested that the $\Phi^4$ theory is trivial claiming it  in $d\neq 4$ \cite{Frohlich}, \cite{Aizenmann}, \cite{Jaffe}, in computer simulations \cite{Callaway} or for $O(N)$ symmetric model \cite{Bardeen}. In \cite{Frasca}, \cite{Jora}  it was proved, based on non perturbative methods,  that the $\Phi^4$ theory is trivial for a large bare coupling constant.

In the present work we will give a proof of triviality of the $\Phi^4$ theory valid for any value of the the bare coupling constant.  Thus we will show that the all order propagator of the theory is that of the free Lagrangian:
\begin{eqnarray}
\langle\Phi(x)\Phi(y)\rangle=\int \frac{d^4p}{(2\pi)^4}e^{-ip(x-y)}\frac{1}{p^2-m_0^2}.
\label{trtr657}
\end{eqnarray}
This definition is more restrictive than that suggested in \cite{Frasca} but it is particular case of it.

 Consider a simple $\Phi^4$ theory without spontaneous symmetry breaking with the Lagrangian:
 \begin{eqnarray}
 {\cal L}=\frac{1}{2}\partial^{\mu}\Phi\partial_{\mu}\Phi-\frac{1}{2}m_0^2\Phi^2-\frac{\lambda}{4}\Phi^4.
 \label{lagr4567}
 \end{eqnarray}
 The all order two point function has the well known expression:
 \begin{eqnarray}
 \langle [\Phi(x)\Phi(y)]\rangle=
 \int \frac{d^4p}{(2\pi)^4}e^{-ip(x-y)}\frac{i}{p^2-m_0^2-M(p^2)},
 \label{rez435}
 \end{eqnarray}
 where $M(p^2)$ is the all order correction to the scalar mass.
Here it is assumed that a cut-off procedure is used and no attempt at renormalization is made. Let us write the two point function explicitly in the Fourier space:
\begin{eqnarray}
\langle [\Phi(x)\Phi(y)]\rangle=
\int \frac{d^4p}{(2\pi)^4}\frac{d^4q}{(2\pi)^4}e^{-ipx}e^{-ipy}\langle\Phi(p)\Phi(q)\rangle.
\label{four5467}
\end{eqnarray}
Note that the quantity $\langle \Phi(p)\Phi(q) \rangle$ is given in the path integral formalism by:
\begin{eqnarray}
\langle\Phi(p)\Phi(q)\rangle=
\frac{\int \prod_k d\Phi(k)\Phi(p)\Phi(q)\exp[i \int d^4x {\cal L}]}{\int \prod_k d\Phi(k)\exp[i \int d^4x {\cal L}]}.
\label{ev5534567}
\end{eqnarray}

 Assume that instead the quantity in Eq. (\ref{four5467}) we consider in the Fourier space the function:
 \begin{eqnarray}
 I_{(x-y)}=\int \frac{d^4p}{(2\pi)^4}e^{-ip(x-y)}\langle\Phi(p)\Phi(-p)\rangle.
 \label{func546789}
 \end{eqnarray}
We need to find the significance of this function  in the coordinate space:
\begin{eqnarray}
&&\int \frac{d^4p}{(2\pi)^4}e^{-ip(x-y)}\langle\Phi(p)\Phi(-p)\rangle=
\nonumber\\
&&\int \frac{d^4p}{(2\pi)^4}\int d^4 z_1 d^4 z_2\langle \Phi(z_1)\Phi(z_2) \rangle e^{ipz_1}e^{-ipz_2}e^{-ip(x-y)}=
\nonumber\\
&&\int d^4 z_1 d^4 z_2\delta(z_1-z_2-x+y)\langle \Phi(z_1)\Phi(z_2) \rangle=\int d^4 z_2\langle \Phi(z_2+x-y)\Phi(z_2)\rangle.
\label{rez43567}
\end{eqnarray}
Knowing that,
\begin{eqnarray}
\langle [\Phi(z_2+x-y)\Phi(z_2)]\rangle=\int \frac{d^4p}{(2\pi)^4}e^{-ip(x-y)}\frac{i}{p^2-m_0^2-M^2(p^2)}= \langle [\Phi(x)\Phi(y)]\rangle
\label{equal7689}
\end{eqnarray}
we obtain the following relation between $I_{(x-y)}$ and the two point function:
\begin{eqnarray}
 I_{(x-y)}=\int d^4 z_2\langle [\Phi(x)\Phi(y)]\rangle=\int d^4 z_2 \int \frac{d^4p}{(2\pi)^4}e^{-ip(x-y)}\frac{i}{p^2-m_0^2-M(p^2)}.
 \label{rez615789}
\end{eqnarray}
Note that $\int d^4 z_2$ is independent of the rest of the expression and if one considers the functional integration over a lattice with the volume V one can simply write $\int d^4 z_2=V$.

Now we shall calculate the function $I_{(x-y)}$ in the path integral formalism.

First we write the action for the Lagrangian in Eq. (\ref{lagr4567}) in the Fourier space.
\begin{eqnarray}
\int d^4 x {\cal L}=\frac{1}{V}\sum_p \frac{1}{2}(p^2-m_0^2)\Phi(p)\Phi(-p)-\frac{\lambda}{4}\frac{1}{V^3}\sum_{k,n,m}\Phi(p_n)\Phi(p_m)\Phi(p_k)\Phi(-p_k-p_m-p_n).
\label{rez3245}
\end{eqnarray}
Then we compute in the path integral formalism:
\begin{eqnarray}
\langle\Phi(p)\Phi(-p)\rangle=
\frac{\int \prod_k d\Phi(k)\Phi(p)\Phi(-p)\exp[i \int d^4x {\cal L}]}{\int \prod_k d\Phi(k)\exp[i \int d^4x {\cal L}]}=\frac{V}{i}\frac{\delta Z}{\delta p^2}\frac{1}{Z},
\label{ev34567}
\end{eqnarray}
where Z is the zero current partiton function:
\begin{eqnarray}
Z=\int \prod_k d \Phi(k) \exp[i \int d^4x {\cal L}].
\label{part4567}
\end{eqnarray}
Here we used the fact that the following relations hold:
\begin{eqnarray}
\int \prod_k d\Phi(k)\Phi(p)\Phi(-p)\exp[i \int d^4x {\cal L}]=\int \prod_k {\rm Re}\Phi(k) {\rm Im}\Phi(k)\exp[\frac{i}{V}\sum_{p^0>0}(p^2-m_0^2)[({\rm Re}\Phi(p))^2+({\rm Im}\Phi(p))^2]+...],
\label{wri777}
\end{eqnarray}
and,
\begin{eqnarray}
\langle\Phi(p)\Phi(-p)\rangle=\langle [({\rm Re}\Phi(p))^2+({\rm Im}\Phi(p))^2]\rangle.
\label{rez32455}
\end{eqnarray}

\section{Partition function}

Before going further we need to establish some facts about the zero current partition function Z. For that we write explicitly:
\begin{eqnarray}
Z=\int \prod_k d \Phi(k)\exp[\frac{i}{2V}\sum_p(p^2-m_0^2)\Phi(p)\Phi(-p)][1-i\frac{\lambda}{4}\frac{1}{V^3}\sum_{k,n,m}\Phi(p_n)\Phi(p_m)\Phi(p_k)\Phi(-p_k-p_m-p_n)+...],
\label{part111}
\end{eqnarray}
where an infinite expansion in $\lambda$ (the interaction term) is considered. First we note that any term in the expansion gives contributions only if it contains pairs of the type $\Phi(k)\Phi(-k)$ and any such pair can be written as:
\begin{eqnarray}
\Phi(k)\Phi(-k)=\frac{V}{i}\frac{\delta  \exp[i\frac{1}{2V}\sum_p(p^2-m_0^2)\Phi(p)\Phi(-p)]}{\delta k^2}.
\label{rez32456}
\end{eqnarray}
Then one can write:
\begin{eqnarray}
&&Z=\int \prod_k d \Phi(k)\exp[\frac{i}{2V}\sum_p(p^2-m_0^2)\Phi(p)\Phi(-p)][1-i\frac{3\lambda}{4}\frac{1}{V^2}\sum_{k,n}\Phi(p_n)\Phi(-p_n)\Phi(p_k)\Phi(-p_k)+...]=
\nonumber\\
&&=\int \prod_k d \Phi(k)[1-i\frac{3\lambda}{4}\frac{1}{V^2}\frac{V^2}{i^2}\sum_{k,n}\frac{\delta}{\delta p_n^2}\frac{\delta}{\delta p_k^2}+...]\exp[\frac{i}{2V}\sum_p(p^2-m_0^2)\Phi(p)\Phi(-p)]
\nonumber\\
&&=[1-i\frac{3\lambda}{4}\frac{1}{V^2}\frac{V^2}{i^2}\sum_{k,n}\frac{\delta}{\delta p_n^2}\frac{\delta}{\delta p_k^2}+...]\frac{1}{\det[\frac{i}{V}(p_m^2-m_0^2)]}.
\label{calc657890}
\end{eqnarray}
Since the determinant is diagonal one obtains:
\begin{eqnarray}
&&\frac{V}{i}\frac{\delta }{\delta p^2}\frac{1}{p^2-m_0^2}=iV\frac{1}{p^2-m_0^2}\frac{1}{p^2-m_0^2}
\nonumber\\
&&\frac{V}{i}\frac{\delta }{\delta p^2}\frac{1}{\det[\frac{i}{V}(p_m^2-m_0^2)]}=iV\frac{1}{p^2-m_0^2}\frac{1}{\det[\frac{i}{V}(p_m^2-m_0^2)]}
\label{calc65789}
\end{eqnarray}
Although the procedure is more intricate this type of results are valid for any terms in the expansion of the interaction Lagrangian. Noting from the last line in Eq. (\ref{calc657890}) that these terms are summed (or integrated) over the momenta  one can conclude that besides the determinant that appears in the expression of the partition function there is no other contribution that depends on individual momenta as in all other contributions the momenta are summed over. Thus one can determine that the all orders partition function has the expression:
\begin{eqnarray}
Z=\frac{1}{\det_{p^0>0}[\frac{i}{V}(p_m^2-m_0^2)]}\times {\rm const},
\label{final54678}
\end{eqnarray}
where the factor $const$ depends on the regularization procedure but it is independent on the individual momenta.

\section{Conclusion}

From Eqs. (\ref{ev34567}) and (\ref{final54678}) we then determine:
\begin{eqnarray}
\langle\Phi(p)\Phi(-p)\rangle=\frac{V}{i}\frac{\delta Z}{\delta p^2}\frac{1}{Z}=\frac{iV}{p^2-m_0^2}.
\label{ev345678}
\end{eqnarray}
Note that this is an all orders result.

Furthermore Eqs. (\ref{func546789}), (\ref{rez615789}) and (\ref{ev345678}) lead to:
 \begin{eqnarray}
 &&I_{(x-y)}=\int \frac{d^4p}{(2\pi)^4}e^{-ip(x-y)}\langle\Phi(p)\Phi(-p)\rangle=
 \nonumber\\
 &&=\int \frac{d^4p}{(2\pi)^4}e^{-ip(x-y)}\frac{iV}{p^2-m_0^2}=\int d^4 z_2 \int \frac{d^4p}{(2\pi)^4}e^{-ip(x-y)}\frac{i}{p^2-m_0^2-M(p^2)}
 \label{func11546789}
 \end{eqnarray}
Considering the  assumption that we work on a lattice with the volume V ($\int d^4 z_2=V$) Eq. (\ref{func11546789}) yields the all orders relation:
\begin{eqnarray}
\frac{i}{p^2-m_0^2}=\frac{i}{p^2-m_0^2-M(p^2)},
\label{rez32456}
\end{eqnarray}
which shows that the all order $\Phi^4$ theory is trivial in the sense that the nonperturbative complete two point function receives no corrections from the $\lambda$ term or the all order $M(p^2)=0$. Note that this result does not contradict the all order result for the mass corrections computed in \cite{Jora} since $m^2=m_0^2$ where $m$ is the physical mass is also a solution of the recurrence relations found there.

Thus we completed the proof that the $\Phi^4$ theory is trivial in the sense that it behaves like a free non interacting theory. No assumption about the value of $\lambda$ is made so this result is valid for both small couplings and large couplings regimes.
The method in this work needs adjustments in order to be applicable for a scalar with spontaneous symmetry breaking or for other cases in quantum field theories.

\section*{Acknowledgments} \vskip -.5cm

The work of R. J. was supported by a grant of the Ministry of National Education, CNCS-UEFISCDI, project number PN-II-ID-PCE-2012-4-0078.

\end{document}